\documentclass[12pt,preprint]{aastex}

\newcommand{\fwhm}{{\small FWHM}}

\newcommand{\chandra}{{\it Chandra}}

\newcommand{\hetgs}{{\small HETGS}}

\newcommand{\x}{X-ray}

\newcommand{\kms}{km~s$^{-1}$}
\newcommand{\etc}{$\eta$~Car}
\newcommand{\msol}{$M_{\odot}$}
\newcommand{\msolyr}{$M_{\odot}$~yr$^{-1}$}
\newcommand{\etal}{et~al.}

\begin{document}

\title{A high-velocity transient outflow in $\eta$ Carinae}
\author{Ehud Behar\altaffilmark{1}, Raanan Nordon\altaffilmark{1},
 and Noam Soker\altaffilmark{1}}

\altaffiltext{1}{Department of Physics,
                 Technion, Haifa 32000, Israel.
                 behar@physics.technion.ac.il, nordon@tx.technion.ac.il,
                 soker@physics.technion.ac.il.}
%\received{} \revised{7/1/04} \accepted{}

\shorttitle{transient outflow in $\eta$ Carinae}
\shortauthors{Behar et al.}

\begin{abstract}
We analyze velocity profiles of the \x\ spectral lines emitted by
the $\eta$~Carinae (\etc) stellar binary at four epochs, just
before the \x\ minimum (associated with periastron) and more than
two years before the minimum ($\sim$ apastron). The profiles are
nicely resolved by the \hetgs\ spectrometer on board \chandra. Far
from periastron, we find symmetrical lines that are more or less
centered at zero velocity. Closer to periastron, the lines
broaden, shift towards the blue, and become visibly asymmetric.
While the quiescent \x\ emission and slight ($<200$~\kms) centroid
shifts can be ascribed to the ordinary continuous binary wind
interaction and to the orbital velocity of the secondary star, the
observed high-velocity emission up to $\sim$2,000 \kms\ and the
abrupt flares during which it occurs can not. This leads us to
interpret the high-velocity flaring emission as due to a fast
collimated outflow of ionized gas.

\end{abstract}


\section{INTRODUCTION}
\label{sec:intro}

Despite being studied for many years and in all wavebands, the
$\eta$ Carinae system (\etc), best known for its spectacular
nebula \citep{morse98} that can be tracked back to the Great
Eruption of 1840 \citep{davidson97} is still much of a mystery. It
is now believed to be a stellar binary system with a period of
5.54 years \citep{damineli96}. The primary star is one of the
brightest and most massive ($\sim$ 120~\msol) stars in our galaxy.
%The system is currently in a phase of immense mass ejection.
Although the \x s detected from \etc\ can be largely explained by
the collision of the two stellar winds \citep{corcoran01,
pittard02},
%, akashi06, hamaguchi07},
the $\sim$~70 day \x\ minimum that is periodically observed around
periastron is still not well-understood.

Currently, the mass loss rate of the primary exceeds
$10^{-4}$\msolyr\ \citep{smith03} and that of the secondary is an
order of magnitude lower \citep{pittard02}. The wind of the
secondary is much faster \citep[few 1,000 \kms,][]{pittard02} than
that of the primary \citep[$\sim$~600 \kms\ and depending on
latitude,][]{smith03}. The \x\ flux between 2 and 10~keV during
the past 12 years is normally at a level of 5$\times 10^{-11}$
erg~s$^{-1}$~cm$^{-2}$, as expected from the collision of the two
stellar winds \citep{usov92, corcoran01, pittard02, akashi06}. The
\x\ light curve is roughly constant throughout most of the orbit,
but rises gradually by a factor of 3--4 upon approach to
periastron at which time short flares appear \citep{corcoran97},
before it drops sharply and stays in an \x\ low state ($\sim$~10\%
of normal brightness) for approximately 70~days. The full X-ray
light curve can be found in \citet{corcoran05}. Although
absorption by the dense primary wind plays a certain role in
attenuating the X-ray emission from \etc , both spectral and
temporal arguments have been put forward to reject absorption or
an eclipse as the prime reason for this persisting low state
\citep{hamaguchi04, soker05, akashi06}, which is unique to \etc.
It has been proposed that if the binary separation during
periastron is small enough, the massive primary wind could smother
the secondary wind and shut down the X-ray emission
\citep{soker05, akashi06}. Recently, a detailed study of the
broad-band \x\ behavior around the 2003 periastron passage was
conducted by \citet{hamaguchi07}, who claim the unchanged
temperature during the \x\ minimum argues against accretion.
However, if the \x\ emission during the minimum is due to the
secondary's residual fast polar wind \citep{akashi06}, it would
indeed result in unchanged \x\ temperatures.

In this work, we wish to add to the temporal aspects of the \etc\
tale a high-resolution \x\ spectroscopic dimension. This is
achieved by using five deep archival exposures of \etc\ with the
High Energy Grating Spectrometer (\hetgs) on board the \chandra\
\x\ Observatory (CXO), only the first of which has been published
\citep{corcoran01,corcoran01b}. Theoretical variations of spectral
line profiles in binary colliding wind systems have been modeled
by \citet{henley03}. The varying line profiles presented here for
\etc, however, do not follow these models, at least not in a
straightforward way, as we explain in \S 3.

\section{OBSERVATIONS AND RESULTS}

The log of the \chandra\ observations is given in Table~1.   Two
observations (3745 and 3748) coincide with the short intense
flares just before periastron.  The exact orbital parameters of
\etc\ are not well determined, but we assume throughout this paper
a period of 5.54~years (2024~days) and periastron to be at
2003-06-29. The X-ray low state, thus, occurs during $\phi$~= 0 --
0.035. The data were retrieved from the CXO archive and processed
using standard software (CIAO version 3.2.2.). \chandra 's HEG and
MEG (high and medium energy gratings) spectrometers operate
simultaneously in the 3.5 -- 7.5 \AA\ band of interest here.
Aiming at the highest possible kinematic resolution, the higher
spectral resolving power of the HEG in this band (factor of $\sim$
2) is favored over the slightly higher effective area of the MEG
(factor of $\sim$ 2), especially since the co-added plus and minus
1st order HEG data are already of sufficiently high
signal-to-noise ratio (S/N).

To reveal the dynamics of the \x\ gas, we transferred each
spectral line profile to line-of-sight velocity space ($v_\|$ in
\kms) using the non-relativistic Doppler shift $v_\| = c(\lambda -
\lambda _0)/ \lambda _0$, where $\lambda$ and $\lambda _0$
represent the observed and rest-frame wavelengths of the line,
respectively, and $c$ is the speed of light. Negative and positive
velocities represent approaching and receding gas, respectively.
Fig.~1 shows the broad and variable \x\ line profiles of \etc\
demonstrated on the Si$^{+13}$ Ly$\alpha$ doublet at 6.18~\AA. As
expected, the MEG profile represents a smoothed version of the
higher-resolution HEG profile. The observed line is much broader
than the instrumental line spread function (also shown). Fig.~1
also demonstrates how the line profile varies dramatically from
phase $\phi$~= --0.470 (Obs. 632) to $\phi$~= --0.028 (Obs. 3745).
The early-phase ($\phi$~= --0.470) line is centered around zero
velocity with no significant emission beyond 700~\kms. The line
closer to periastron ($\phi$~= --0.028) features high velocity gas
up to $\sim$2000~\kms.
%A gaussian fit yields a centroid shift of
%--550 $\pm$~90~\kms, although the profile is clearly asymmetric.
%we show a sample spectrum (Obs. 3748 at assumed orbital phase
%--0.006) with the main lines identified to exemplify the
%high-quality line-resolved spectra. The steep decline of the
%continuum towards long wavelengths manifests the photoelectric
%absorption along the line of sight, most likely by the primary
%wind.
%The inset in Fig.~1 highlighting the
%Si lines demonstrates the abnormal line profiles and
%approaching-velocity Doppler shifts observed when the system is
%presumably close to periastron.

In each observation, the kinematic profiles of the bright spectral
lines are fairly similar. This is demonstrated in Fig.~2, where
the four best S/N lines of Obs.~3745 ($\phi$~= --0.028) are shown.
%This is demonstrated in Fig.~2, where we present four bright
%spectral lines from Obs.~3745 ($\phi$~= --0.028).
%The profiles are unambiguously broad and asymmetric with
%approaching velocities reaching 2,000~\kms, but with no
%significant receding emission beyond $\sim$~700~\kms.
In order to elucidate how the line profiles vary as a function of
phase, we co-added the profiles of nine bright lines in each
observation, namely the Ly$\alpha$ (1s-2p), He$\alpha$ resonance
(1s$^2$-1s2p) and He$\alpha$ forbidden (1s$^2$-1s2s) lines of the
Si, S, and Ar ions. The similarity of the profiles of these lines
in a given observation \citep[see also][Figs.
4.9--4.12]{henleyphd} justifies the co-adding procedure, which
yields an average profile with improved S/N. The sufficiently
large separation of the resonance to forbidden lines (3400 -- 4100
\kms ) and the weak intensity of the intercombination lines in the
He-like triplets ensure that the average profiles are not confused
by blends. Individual Fe lines below 2~\AA\ could not be included
in this analysis, despite their high intensity, since the \hetgs\
resolving power decreases strongly at these low wavelengths to the
extent that the He-like Fe line complex is unresolved.  The blue
wing of the He$\alpha$ Fe resonance line is consistent with,
though not resolved as well as, the other line profiles presented
here. The resulting mean profiles have been normalized (except for
Obs. 3747) to facilitate the comparison and are presented in
Fig.~3.

%The observed, mean, unnormalized peak intensities for the assumed
%phases: --0.470, --0.130, --0.028, --0.006, and +0.044 are,
%respectively, 2.0, 3.9, 3.9, 1.9, and 0.3,
%$\times$~10$^{-3}$~ph~s$^{-1}$~cm$^{-2}$~\AA$^{-1}$. However, any
%interpretation of the {\it absolute} line fluxes in terms of the
%orbital parameters would be risky, as the observed variability is
%a result of both abrupt flaring and varying absorption.

A clear trend can be seen in Fig.~3. Far from periastron, the line
profile is relatively symmetric and narrow. Closer to periastron,
bright components of gas moving towards us at velocities as high
as --2,000~\kms\ start to develop. As the system further
approaches periastron, the outflow dominates the line profile to
the point where the bulk of the emission is clearly blue-shifted.
After periastron, the emission lines are considerably weaker. Line
profiles from a binary wind system are not expected to be Gaussian
\citep{henley03}. Nevertheless, in order to get an idea of the
line shifts and widths, we tried to fit Gaussians to the mean
profiles in the --2,500~\kms\ to +1,300~\kms\ range (Fig.~3). The
results are presented in Table~2. For the two late-phase
observations ($\phi$~= --0.028, --0.006) single Gaussians provide
poor fits that can be significantly improved by adding a second
component. This does not imply there are necessarily exactly two
velocity components, but merely that the velocity distribution at
these phases  is more complex than Gaussian. The double-Gaussian
best fits are plotted in Fig.~3. A second isolated peak at
$\phi$~= --0.028 is not as conspicuous as at $\phi$~= --0.006.
Hence the broad high-velocity component and the larger errors at
$\phi$~= --0.028 (Table~2).

%This simply means that the high-velocity gas, clearly
%present both at $\phi$~= --0.028 and at $\phi$~= --0.006 (see
%Fig.~2) is not very well represented by a Gaussian distribution.

%It is hard to rule out high-velocity
%{\it receding} gas before the \x\ minimum, since in agreement with
%\citet{hamaguchi04}, we find that the column density toward the
%\x\ source rises from 5$\times$10$^{22}$~\cmsq\ long before
%periastron to 3$\times$10$^{23}$~\cmsq\ just after it. Emission
%from the far side of the system, thus, could be strongly absorbed.

%as the \x\ absorption is observed to increase
%significantly before the minimum.
%Measuring the absorbing column density, we find that it rises from
%5$\times$10$^{22}$~\cmsq\ before periastron to
%3$\times$10$^{23}$~\cmsq\ right after it, where the emission lines
%are almost completely quenched, and in agreement with previous
%reports \citep{hamaguchi04}.

\section{INTERPRETATION AND DISCUSSION}

The colliding wind binary model, which is generally accepted for
explaining the \x\ emission of \etc, has several clear predictions
regarding the spectral line profiles expected at different orbital
phases.  The Doppler shifts are the result of shocked gas flowing
along the contact discontinuity (CD) surface and away from the
stagnation point (SP). If the shock opening half-angle is between
$45^\circ\ - 90^\circ$, the broadest line profiles are expected
when the system is observed perpendicular to the line connecting
the two stars, as the hottest gas formed around the SP flows
directly towards and away from the observer. In this case, the
lines are centered at zero velocity and are intrinsically
symmetrical, but could be skewed by absorption of the far (red)
emission.  On the other hand, when the system is observed from
behind one of the stars, shifted centroids are expected. If the
system is observed from behind the weaker (stronger) wind,
downstream shocked gas flows along the CD surface in the general
direction of (opposite) the observer, producing blue- (red-)
shifted centroids. The maximum velocities in this case would be
less than those in the first (orthogonal) orientation. For a
complete and detailed calculation of line profiles from colliding
winds see \citet{henley03}. In the following, however, we argue
that the current line profiles observed at the different phases of
the \etc\ orbit are in contradiction, even qualitatively, with the
simple colliding wind scenario.

If the major axis of the \etc\ binary is perpendicular to our line
of sight \citep{smith04}, the broadest profiles are expected at
apastron (see above), but that is when the observed profiles are
narrowest ($\phi$~= --0.470).  Also, with this geometry, the
secondary would have to pass in front of the primary before
periastron \citep[c.f.,][]{smith04} to explain the blue-shifted
centroids at $\phi$~= --0.028, --0.006, but that would produce
symmetric (blue-shifted) profiles as absorption through the
secondary wind is weak, while asymmetric profiles are observed. In
any case, the lower velocities observed near apastron appear to
rule out this orientation if the colliding wind geometry is to
produce the observed \x\ line profiles.

Alternatively, if the projection of the line of sight on the
orbital plane is along the major axis and the viewing angle at
apastron is from behind the secondary \citep{corcoran01}, the
observed blue-shifts at $\phi = -0.028, -0.006$ could be ascribed
to gas flowing from the SP towards the observer and the unobserved
red wing of the line to absorption by the primary wind. However,
according to this scenario, at apastron ($\phi = -0.470$) one
would expect blue-shifted centroids of at least a few 100~\kms\
due to downstream gas (see above), which are not observed. For an
observer above the binary plane ($i < 90^\circ$), as seems to be
the case in \etc\ \citep{corcoran01}, the problem becomes more
severe (higher blue-shift is expected, but not observed). Indeed,
\citet{kashi07} showed that the \x\ line shifts with orbital phase
can not be due to the change in orientation of the colliding winds
region when the inclination of the orbital plane is considered
\citep[c.f.,][]{corcoran07}.

Apparently, the naive wind-collision scenario alone is
insufficient to explain the observed Doppler shifts of the \x\
line profiles of \etc\ at all binary phases. \citet{henleyphd}
compared the same data with line profiles calculated from his
hydrodynamical simulations and reached a similar conclusion. A
recent attempt has been made by \citet{henley07} to propose yet
another orientation.
%He suggested that including the Coriolis effect may bring the models
%to better agreement with observations.
However, we note that the asymmetric line profiles are observed
during strong peaks in the X-ray light curve. These peaks too can
not be accounted for by the simple colliding wind models.
%\citep{pittard02, henley03, corcoran07}, nor by the Coriolis
%force.
As far as we understand, the high \x\ velocities remain
unexplained and we ascribe them to a transient outflow that arises
just before the onset of the \x\ minimum and is coincident with
the \x\ flaring.

\citet{kashi07} suggested orbital motion to be important before
periastron.  This could account for small \x\ line shifts, but not
for the highest velocities observed. Utilizing the two-Gaussian
fits in Table~2, however, we can confront the orbital solution of
\citet{kashi07} with the low-velocity fitted component (column~2
in Table~2).
%The small shifts of the main peaks are of the order
%of the expected orbital motion and of the somewhat ($\sim 20\%$)
%lower velocity of the shocked \x\ gas that is dragged with the
%secondary. In Fig.~4 we compare these model velocities with the
%observed shifts as a function of phase.
We assume an eccentricity of $e=0.9$, stellar masses of $M_1=120
M_\odot$, $M_2=30 M_\odot$, and an observer $37 ^\circ$ above the
orbital plane ($i=53 ^\circ$),
%and behind the secondary at periastron \citep{kashi07}.
whose projection on the orbital plane is exactly behind the
secondary at periastron. This geometry was invoked by
\citet{kashi07} to explain the shifts of the He\,I lines
\citep[but c.f.,][]{corcoran01, henley07}.
%The main \x\ source is believed to be the shocked wind of
%the secondary star.
The calculation shows that the orbital velocity of the \x\ gas is
$\sim 80\%$ that of the secondary. We added --8~\kms\ to the model
for the systemic velocity of \etc\ \citep{smiths04}. The resulting
theoretical velocities of the \x\ gas as a function of phase are
seen in Fig.~4 to formally agree with the observed Doppler shifts
of the low-velocity component, except at $\phi$~= --0.006, where
the model velocity is slightly smaller.
% are compared in the bottom panels of Fig.~4 with the
% observed shifts as a function of phase.
Given the many uncertainties of the binary parameters, this
discrepancy may not be significant.

The high-velocity \x\ emission (column~4 in Table~2), on the other
hand, is not explained by the orbital motion and could be due to a
transient collimated flow ejected from the immediate vicinity of
the binary system. The peaks in the X-ray light curve hint that
the outflow may be in the form of blobs. The unresolved hard \x\
\chandra\ images restrict the outflow to within $\sim
2\times$10$^{16}$~cm of the center (0.5\arcsec , assuming the
distance of 2.3~kpc). The appearance of the same charge states
during all phases implies that the temperatures of the X-ray gas
and line-emitting outflow remain in the range of $kT$~= 2~--
5~keV. Hence, the outflow likely consists of gas shocked by the
collision of the winds (as observed throughout the orbit) and is
boosted near periastron. The widths of the major peaks seen in
Figs.~2 and 3 suggest that the outflow is only moderately
collimated to within $\sim 30^\circ$. The compression of shocked
gas provides a natural explanation for the short intermittent
flares in the \x\ light curve. A collimated outflow from the
secondary has also been suggested to explain the enhanced He~II
$\lambda 4686$ emission before periastron \citep{soker06}.
Interestingly, both proper motion and Doppler shifts have been
measured for high-velocity visible-light knots much further out
from the center \citep{walborn78, meaburn93}. The optical knots
were ejected more than a hundred years ago along the minor axis of
the nebula \citep{meaburn93}. We would expect the present outflow
to be directed along the major axis of the nebula although we have
no pertinent information on its present direction with respect to
the system's geometry. As the massive \etc\ primary is a
short-lived star shedding considerable mass, it may be a supernova
in the making. At the very least, it highlights the next
periastron passage on 2009, January 12 as a faithfully scheduled
experiment for the astrophysically common, but poorly understood
phenomenon of collimated flows.

%If indeed that is the case and if the
%inclination angle of the binary plane to our line of sight is $i=
%42^\circ$ - $53^\circ$ \citep{smith02, kashi07}, the actual
%outflow velocity would be as high as $\sim$2,700~\kms.

%X-ray jets have been observed in accreting stellar systems
%\citep{kastner04}, and also in X-ray binaries with a neutron star
%or black-hole companion (microquasars), but not from evolved hot
%stars. Although not well understood, it is known
%phenomenologically that jets are typically found in accreting
%systems.  The current discovery, therefore, makes for
%circumstantial evidence of accretion in the \etc\ system, and as
%such could validate the scenario proposed in \citet{soker05} and
%in \citet{akashi06}. Interestingly, the wind from the progenitor
%of the well studied supernova (SN) 1987A was strongly asymmetric
%\citep[e.g.,][]{allen89}. Furthermore, the galactic SN remnant
%Cassiopeia A has alpha-element-rich jets as well \citep{hwang04},
%which speculatively, could be the scaled up version of those we
%are seeing in \etc.

%As the massive \etc\ primary is a short-lived star shedding
%considerable mass,
%and as it is now understood that
%core collapse supernovae can not expel more than a few solar masses, implying that
% high-mass stars must shed most of their mass prior to their the explosion
%the winds and fast outflow of \etc\ may be a supernova
%in the making. At the very least, it highlights the next
%periastron passage on 2009, January 12 as a faithfully scheduled
%experiment for the astrophysically common, but poorly understood
%phenomenon of collimated flows.

\acknowledgments This research was supported by grant \#28/03 from
the Israel Science Foundation.
%and by a grant from the Asher Space Research Institute at the Technion.

\clearpage

\begin{deluxetable}{lcccc}
\tablecolumns{6} \tablewidth{0pt} \tablecaption{CXO/HETG
Observation Log \label{tab1}} \tablehead{
   \colhead{CXO archive} &
   \colhead{Start Time} &
   \colhead{Exposure} &
   \colhead{Assumed Orbital} &
   \colhead{2--10~keV Flux} \\
   \colhead{ID \tablenotemark{a}} &
   \colhead{} &
   \colhead{(ks)} &
   \colhead{Phase ($\phi$)\tablenotemark{b}} &
   \colhead{(10$^{-10}$ erg~s$^{-1}$~cm$^{-2}$)}}
\startdata
632 & 2000-11-19 02:46:40 & 90.69 &  --0.470 & 0.50 \\
3749 &  2002-10-16 08:08:49 & 93.96 &  --0.130 & 0.98 \\
3745 &  2003-05-02 11:56:16 & 97.29 &  --0.028 & 2.2 \\
3748 &  2003-06-16 05:35:28 & 100.1 &  --0.006 & 0.97 \\
3747 &  2003-09-26 22:45:53 & 72.16 &  +0.044 & 0.48 \\
\enddata
\tablenotetext{a}{ One additional observation was carried out
during the X-ray minimum (ID 3746 at 2003-07-20 01:46:22) for
which no useful spectrum could be obtained.} \tablenotetext{b}{
Phase is approximated based on the assumption of a 5.54~year
(2024~day) orbit, where phase zero is set at 2003-06-29 and the
X-ray low state occurs during phase 0 -- 0.035.}
\end{deluxetable}

\begin{deluxetable}{cccccc}
\tablecolumns{6} \tablewidth{0pt} \tablecaption{Kinematic
parameters in \kms\ obtained from Gaussian fits to the mean line
profiles presented in Fig.~3 \label{tab2}} \tablehead{
   \colhead{Assumed Phase ($\phi$)} &
   \colhead{$v_1$} &
   \colhead{\fwhm$_1$} &
   \colhead{$v_2$} &
   \colhead{\fwhm$_2$} &
   \colhead{$\chi ^2$ / d.o.f \tablenotemark{a}}  }
\startdata
--0.470 & --25 $\pm$ 25 & 710 $\pm$ 60 & \nodata & \nodata & 0.36 \\
--0.130 & --50 $\pm$ 20 & 730 $\pm$ 40 & \nodata & \nodata & 0.43 \\
--0.028 & --110 $\pm$ 30 & 730 $\pm$ 100 & --830 $\pm$ 110 & 2030 $\pm$ 400 & 0.39 \\
--0.006 & --180 $\pm$ 30 & 620 $\pm$ 60 & --1030 $\pm$ 30 & 840 $\pm$ 90 & 0.68 \\
\enddata
\tablenotetext{a}{ Double-Gaussian fits to the profiles at
$\phi$~= --0.470 and at $\phi$~= --0.130 yield an unconstrained
second component. Single-Gaussian models for the profiles at
$\phi$~= --0.028 and $\phi$~= --0.006 yield visibly inferior fits
with $\chi ^2$ / d.o.f of 1.16 and 1.02, respectively. }

\end{deluxetable}

\clearpage

\begin{figure}
  % Requires \usepackage{graphicx}
  \centering
  \includegraphics[width=10.0cm, angle=0]{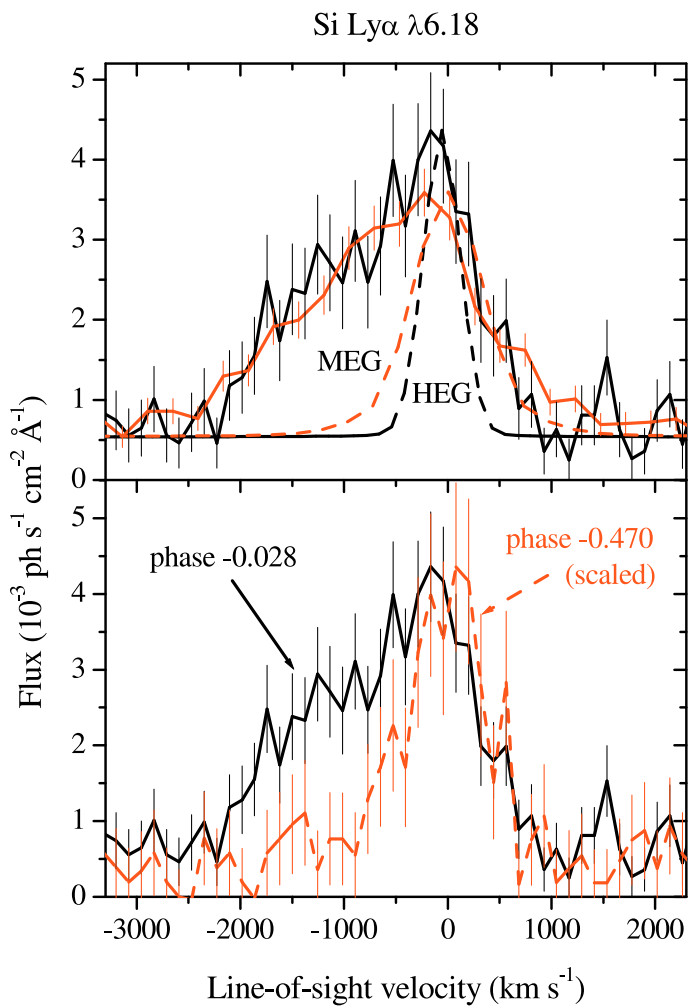}
  \bigskip
  \caption{Velocity profiles of the Si$^{+13}$ Ly$\alpha$ unresolved doublet
(6.180 and 6.186 \AA). {\it Upper panel} shows consistent HEG
(higher spectral resolution) and MEG (smoother) data from Obs.
3745 ($\phi$~= --0.028). The profile is clearly broadened up to
2000~\kms, much beyond the instrument line spread functions
(dashed lines). {\it Bottom panel} shows the HEG profile of Obs.
632 ($\phi$~= --0.470, dashed line) and Obs. 3745 ($\phi$~=
--0.028). The scaled up line from the early phase is rather
symmetrical and much narrower, showing no line emission beyond
700~\kms.}
  \label{f1}
\end{figure}

%\begin{figure}
%\centerline{\includegraphics[width=8.5cm,angle=-90]{spectrum.ps}}
% \caption{\x\ spectrum of \etc\ obtained during Obs.~3748 (assumed phase --0.006)
%  by co-adding plus and minus 1st orders of the HEG spectrometer.
%  The brightest H-like (Ly$\alpha$) and He-like (He$\alpha$, resonance and forbidden) observed lines are indicated at their rest frame.
%  The strong Fe line complex emerges below 2~\AA.  In the inset, a blow-up of the Si lines is given, where the broadening
%  of the lines towards shorter wavelengths can be seen.
%  }
%\label{f1}
%\end{figure}
%

%\clearpage
%
%\begin{figure}
%\centerline{\includegraphics[width=10cm,angle=0]{f2.ps}}
%  \caption{Profiles of four bright spectral lines from Obs.~3745 ($\phi$=~--0.028).
%  The consistent asymmetric profiles with blueshifts up to $\sim\ -2,000$~\kms\
%  are clearly seen.}
%   \label{f2}
%\end{figure}

\clearpage

%\begin{figure}
%  % Requires \usepackage{graphicx}
%  \centering
%  \includegraphics[width=13.0cm, angle=-90]{fig3.ps}
%  \bigskip
%  \caption{Mean spectral line profiles in velocity space for the five CXO/HETG observations of \etc.
%  See Table~1 for details. Errors are the mean 1$\sigma$.  The profiles demonstrate a slight shift of --250~\kms\ and the
%  conspicuous rise of emission from hot, fast outflowing gas at velocities as high as --2,000~\kms\
%  in the two pre-minimum observations.  In the post-minimum spectrum, the emission lines are strongly absorbed.}
%  \label{f3}
%\end{figure}

\begin{figure}
\centerline{\includegraphics[width=10cm,angle=0]{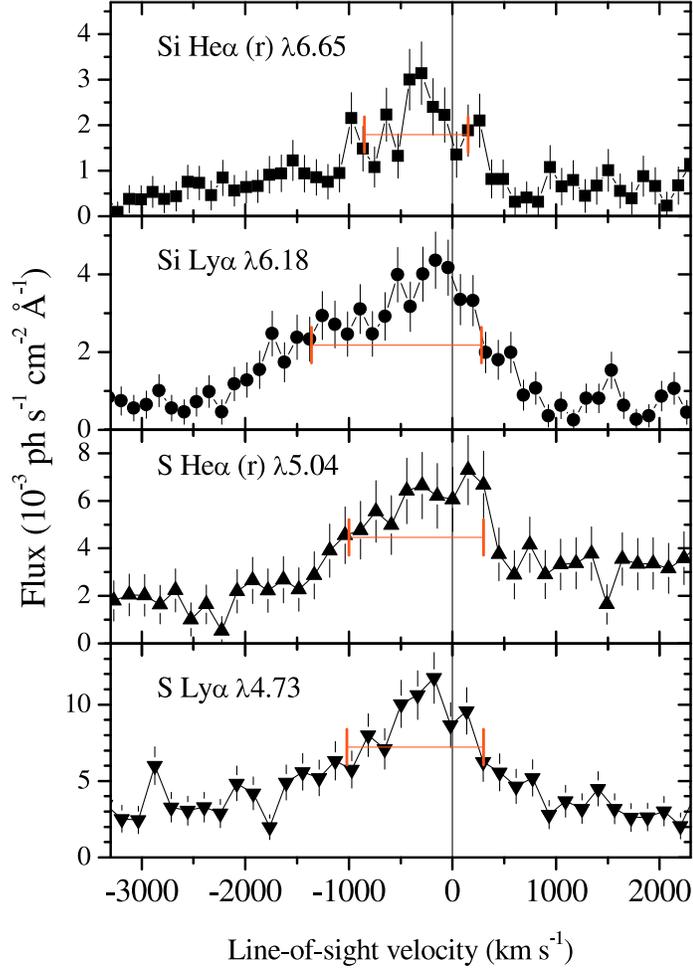}}
  \caption{Profiles of four bright spectral lines from Obs.~3745 ($\phi$~= --0.028) showing
  asymmetric profiles with blueshifts of up to $\sim\ -2,000$~\kms.
  Horizontal bars represent the shift and \fwhm\ Gaussian-fitting results by
  \citet{henleyphd} for these lines, which are not inconsistent with the present profiles.}
   \label{f2}
\end{figure}

\clearpage

\begin{figure}
  % Requires \usepackage{graphicx}
  \centering
  \includegraphics[width=10.0cm, angle=-90]{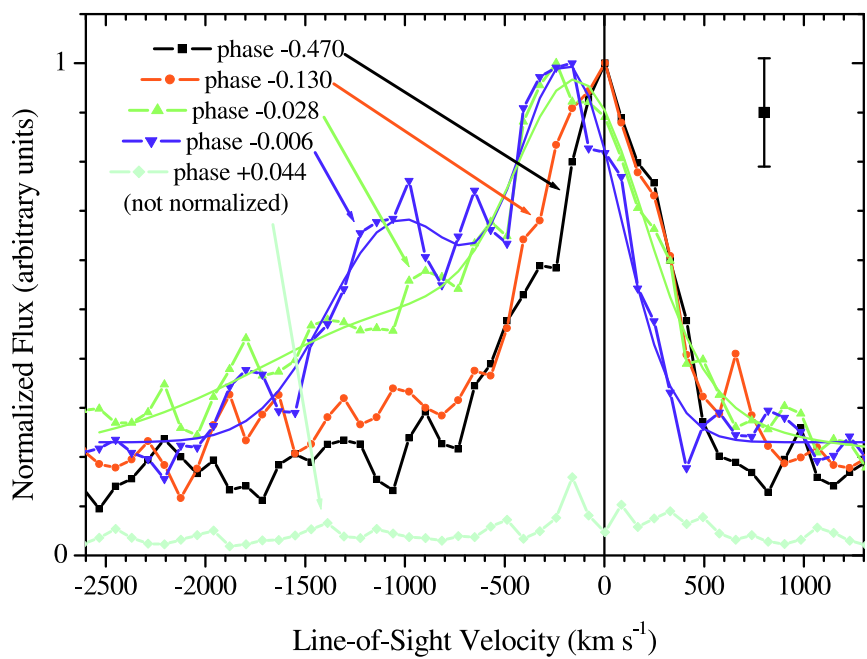}
  \bigskip
  \caption{Mean, normalized, velocity profiles constructed from nine different
  spectral lines for the five observations of \etc.
  Only the profile of Obs.~3747 ($\phi$~= +0.044) during which \etc\ was strongly
  absorbed is not normalized and plotted at its correct scale with respect to Obs. 632
 ($\phi$~= --0.470). Typical 1$\sigma$ errors on each data point are 10 -- 15~\%
 (top right hand side). The double-Gaussian fits to the $\phi$~= --0.028 and $\phi$~=
 --0.006 profiles are also plotted (see Table~2).
%  The profiles demonstrate a slight centroid shift of --250~\kms\ and the
%  conspicuous rise of a fast outflowing emission component at velocities as high as
%  --2,000~\kms\ in the two pre-minimum observations.
}
  \label{f3}
\end{figure}

\clearpage

\begin{figure}
  % Requires \usepackage{graphicx}
  \centering
  \includegraphics[height=11.0cm,angle=0]{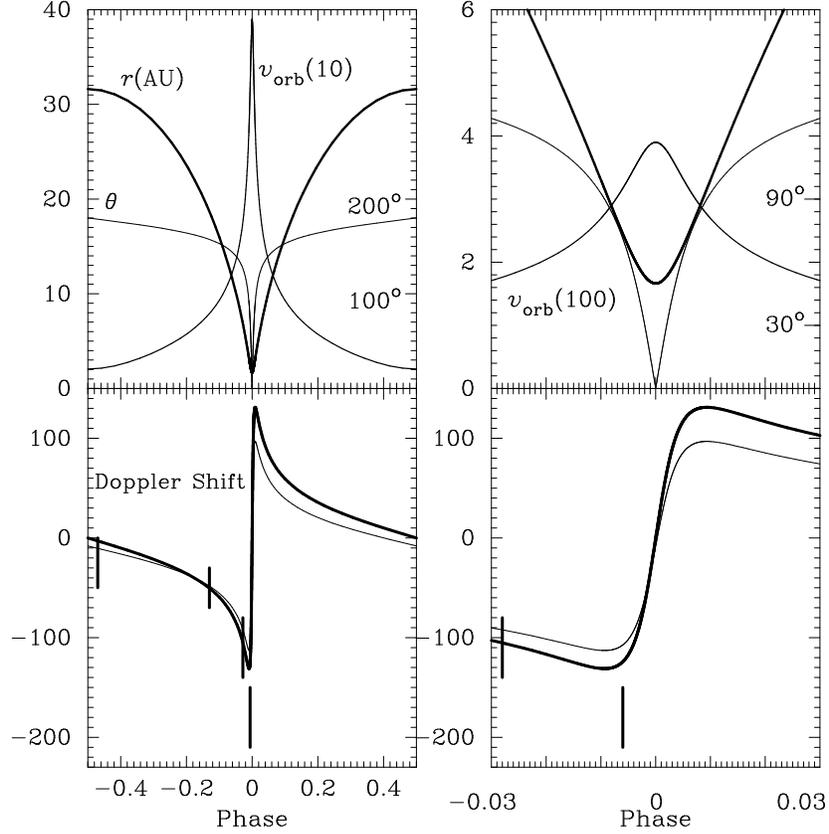}
  \bigskip
  \caption{Several theoretical variables as a function of phase
  assuming the orbital parameters in the text, throughout the orbit
  (left column) and close to periastron (right column).
  Top row: Orbital separation (in AU) and relative orbital speed
  of the two stars (in units of 10 \kms\ on the left and
  100 \kms\ on the right). The angle $\theta$ is the relative
  direction of the two stars as measured from periastron (scale in degrees on the right).
  Bottom row: Projected velocity (in \kms) on the line of sight, of secondary star (thick line)
  and of the \x\ gas (thin line).
  Vertical lines mark the measured Doppler shifts of the slow component
  in Table~2.}
  \label{f4}
\end{figure}

\end{document}